# Electronic correlation strength of inorganic electrides from first principles

Shu Kanno[1], Tomofumi Tada[1,*], Takeru Utsumi[1], Kazuma Nakamura[2], Hideo Hosono[1,†]

[1] *Materials Research Center for Element Strategy, Tokyo Institute of Technology, 4259 Nagatsuta, Midori-ku, Yokohama 226-8503, Japan*

[2] *Quantum Physics Section, Department of Basic Sciences, Kyushu Institute of Technology, 1-1 Sensui-cho, Tobata, Kitakyushu, Fukuoka 804-8550, Japan*

* **tada.t.ae@m.titech.ac.jp**

† **hosono@mces.titech.ac.jp**

## Abstract

We present a systematic study clarifying an electronic correlation trend of electrides from first principles. By using the maximally localized Wannier function and the constrained random phase approximation, we calculated the electronic correlation strength $(U - U_{nn})/|t|$ of 19 inorganic electrides, where $U$, $U_{nn}$, and $t$ are the effective onsite Coulomb repulsion, nearest-neighbor Coulomb repulsion, and the nearest-neighbor transfer integrals, respectively. The electronic correlation was found to be highly correlated with the dimensionality of the Wannier-function network of anionic electrons in electrides; the correlation strength varies in the order 0D >> 1D > 2D ~ 3D, showing good correspondence with experimental trends, and exceeds 10 (a measure for the emergence of exotic properties) in all the 0D systems and some of the 1D materials. We also found that the electronic correlation depends on the cation species surrounding the anionic electrons; in the 1D electrides, the electronic correlation becomes stronger for cationic walls consisting of $Ca^{2+}$, $Sr^{2+}$, and $Ba^{2+}$ in this order, and the correlation strength exceeds 10 for $Ba_5As_3$. The theoretical results indicate that 0- and 1-dimensional electrides will be new research targets for studies on strongly correlated electron systems.

# I. INTRODUCTION

Strongly correlated electron systems, represented by Mott insulators, have attracted attention as a platform for the emergence of exotic properties such as high-Tc superconductivity [1,2]. In general, strongly correlated electron systems are focused on *d*- and *f*-electron systems, and "electronic correlation strength" has been introduced to represent the strength of the electronic correlation [3]. For example, the electronic correlation strength can be a measure of the possibility of the metal-insulator transition in transition metal compounds; the strength of NiO is larger than that of TiO by a factor of four, and the former becomes an insulator with large energy band splitting and the latter a metal [4]. Recently, correlated electron behavior has been observed in a novel materials group, *electrides*, in which *s*-electrons confined in sub-nanometer-sized spaces with different shapes (voids, channels, and interlayer spaces) in intrinsic crystal structures behave like anions (the confined electron is called an anionic electron) [5–13].

The first confirmed electride is the organic crystal $Cs^+$(18-crown-6)$_2$:$e^-$ synthesized by Ellaboudy *et al.* in 1983 [5]. However, the organic electride is so chemically and thermally unstable that research on its properties remains unveiled. An inorganic electride that is superior to organic electrides in terms of stability was synthesized by Matsuishi *et al.* in 2003 as $[Ca_{24}Al_{28}O_{64}]^{4+}(4e^-)$ (hereafter referred to as C12A7:$e^-$) [6]. C12A7:$e^-$ has *s*-electrons confined in crystallographic cages composed of $Ca^{2+}$, $Al^{3+}$, and $O^{2-}$. C12A7:$e^-$ shows a metal-insulator transition when the electron concentration exceeds ~$2 \times 10^{21}$ $cm^{-3}$ [14], and the metallic samples exhibit BCS-type superconductivity [15]. In addition to C12A7:$e^-$, various types of inorganic electrides have been discovered. For example, $Ca_2N$ is a two-dimensional electride in which anionic electrons are sandwiched by cationic slabs $[Ca_2N]^+$ and behave as a two-dimensional (2D) electron gas [7]. Another interesting example is $Sr_5P_3$, a one-dimensional (1D) electride showing Mott insulating behavior; the anionic electrons are confined in channels with $Sr^{2+}$ walls [11]. Thus, electrides are a group of materials that exhibit a variety of properties ranging from insulators to metals. Although the electronic correlation in electrides has been discussed individually [11,12,16–18], a systematic analysis of the electronic correlation of electrides has not been conducted. It is of interest to know the electronic correlation strength for *s*-electrons confined in crystallographic sub-nanometer-sized spaces to gain a comprehensive understanding of the electronic interaction in condensed matter.

In this study, we calculate the electronic correlation strength of inorganic electrides by using first-principles electronic structure calculations. The target systems are 19 inorganic electrides and are classified in terms of the Wannier function framework of anionic electrons, creating an easy-to-understand classification as zero-dimensional (0D), 1D, 2D, and three-dimensional (3D) electrides. Based on this classification, we will show the correspondence between the electronic correlation strength and the electron confinement spaces.

## II. METHODS

The electronic correlation strength is calculated as $(U - U_{nn})/|t|$, where $U$ is the effective on-site Coulomb repulsion, $U_{nn}$ is the effective nearest-neighbor Coulomb repulsion, and $|t|$ is the absolute value of the nearest-neighbor transfer integral. The electronic correlation strength can be used to investigate the trend in the electronic correlation of materials. When the kinetic energy is significantly larger than the potential energy, i.e., $(U - U_{nn})/|t| \ll 1$, the electrons are itinerant throughout the system (i.e., metal). In contrast, when $(U - U_{nn})/|t| \gg 1$, electrons tend to be localized and dominated by the potential energy (i.e., the system becomes a Mott insulator involving magnetic ordering). In the intermediate region, the kinetic energy and potential energy of electrons compete with each other, leading to interesting phenomena. For example, the electronic correlation strengths of $3d$-cuprates, the parent materials of high-Tc superconductors, are known to be ~ 7 [19,20].

In the present study, we calculate the electronic correlation strength of various electrides and investigate the theoretical trend for the electronic correlation of electrides. For this purpose, we calculate the physical parameters $|t|$, $U$, and $U_{nn}$ from first principles based on the procedure given in Fig. 1(a). We demonstrate a concrete procedure for Na-electro-sodalite $Na_8(AlSiO_4)_6$, having well bounded anionic electrons in nanospaces.

In Step I, we perform a density-functional band-structure calculation. The resulting eigenvalues and eigenstates of the Kohn-Sham Hamiltonian are used for the subsequent maximally localized Wannier function calculations (Step II). To obtain the Wannier function for the anionic electron orbital, we set an energy window including the target bands for the Wannier functions, which is set to 3.5 - 4.8 eV in this example. We use an $s$-type Gaussian function centered at the nanospace as the initial guess for the Wannier function calculations. With these settings, the Wannier functions are obtained by minimizing the spread functional [21] and/or the spillage functional for the band-entangled situation [22]. The transfer parameter is determined from matrix elements of the Kohn-Sham Hamiltonian in the Wannier functions as

$$t_{ij} = \int_V \Psi_i(\boldsymbol{r})^* H_0 \Psi_j(\boldsymbol{r})\, d\boldsymbol{r}, \tag{1}$$

where $H_0$ is the Kohn-Sham Hamiltonian, $\Psi_i(\boldsymbol{r})$ is the $i$ th Wannier function, and the integral is taken for the crystal volume $V$. The nearest-neighbor transfer integral $t$ is determined as the matrix element between the nearest-neighbor pair of Wannier functions.

The upper panel in Fig. 1(b) compares the original density-functional band structure (red solid curves) with the Wannier interpolated band (blue crosses) of Na-electro-sodalite, from which we confirm good agreement between the two bands. The lower panel in Fig. 1(b) shows the calculated real-space Wannier function, from which the well-localized character of the sodalite Wannier function in the nanospace is confirmed. The calculated nearest transfer energy $|t|$ is 55 meV.

In Step III, we calculate the effective on-site $U$ and nearest-neighbor $U_{nn}$ Coulomb repulsions with the Wannier functions. The effective direct Coulomb integral is defined as

$$U_{ij} = \iint_V W(\boldsymbol{r}, \boldsymbol{r}')|\Psi_i(\boldsymbol{r})|^2 |\Psi_j(\boldsymbol{r}')|^2 \, d\boldsymbol{r} d\boldsymbol{r}', \qquad (2)$$

where $W(\mathbf{r}, \mathbf{r}')$ is the screened Coulomb interaction based on the constrained random phase approximation (cRPA) [1,23–26]. $U$ is evaluated as the average of the diagonal components of $U_{ij}$. In the sodalite, the values of $U$ and $U_{nn}$ are 2.86 eV and 0.66 eV, respectively, and thus $(U - U_{nn})/|t|$ is obtained as 39.8. Since the value of the strength is quite large, the sodalite is classified as a strongly correlated material [27]. We performed the calculations for $U$, $U_{nn}$, $|t|$, and $(U - U_{nn})/|t|$ of 19 electrides.

As a preprocessing before Step I, full optimization of the crystal structures of the electrides was carried out using VASP [28,29], and for the relaxed structures we performed xTAPP band calculations [30,31] to generate inputs for RESPACK [24,32–36] that calculates the maximally localized Wannier function and the cRPA screened Coulomb interaction. We used the Perdew-Burke-Ernzerhof type [37] for the exchange-correlation functional in both the VASP and xTAPP calculations. For core electrons, the projector augmented wave (PAW) method in VASP and the norm-conserving pseudopotential in xTAPP were adopted. The wave function cut-off, dielectric function cut-off, *k*-point grid, and the number of bands used for each material were determined so that the differences in $U$ converged within 0.1 eV (The values of each parameter are listed in Table A1 of supplemental material A).

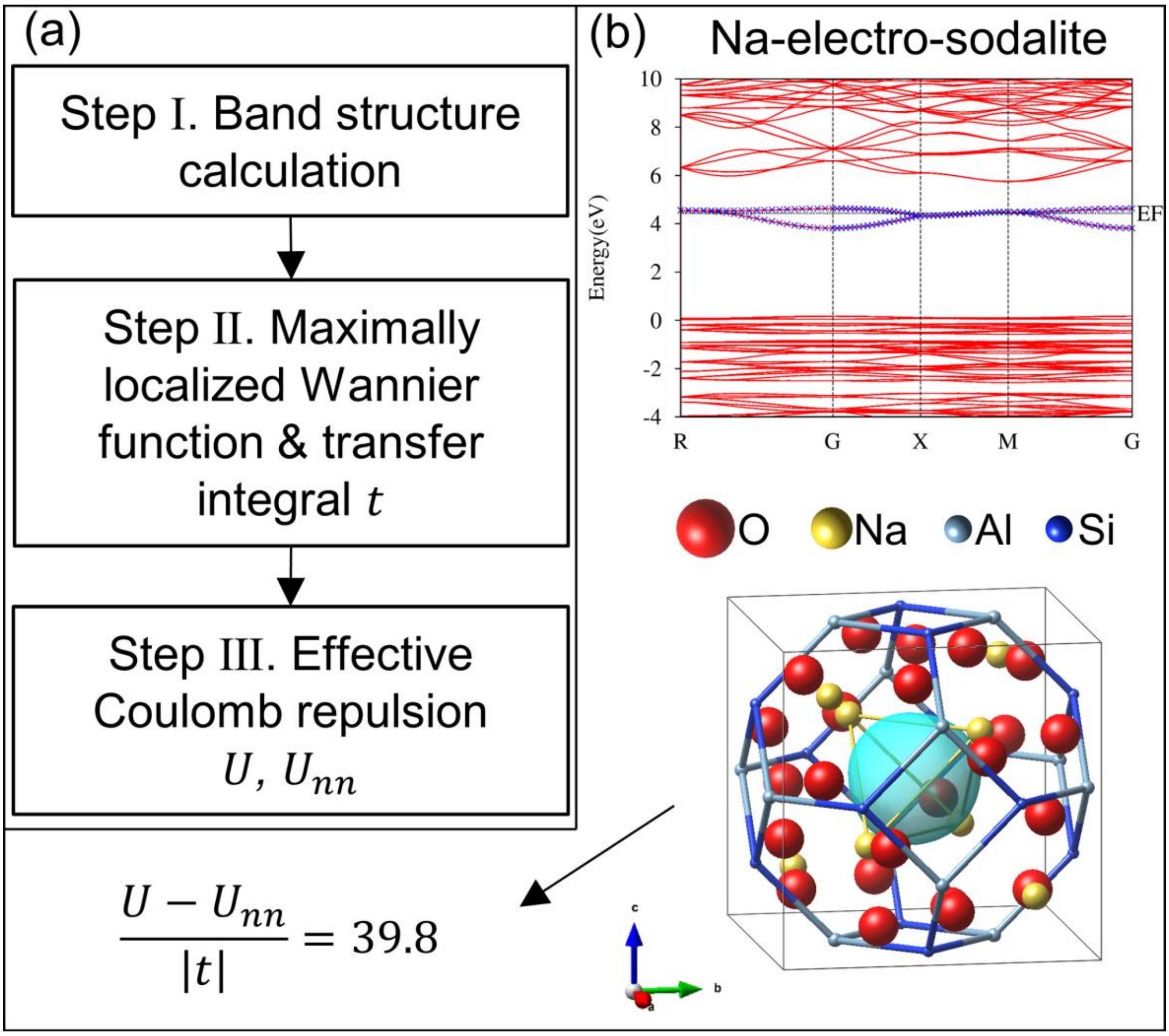

FIG. 1. Computational flow of electronic structure calculations to derive effective parameters and the electronic correlation strength of anionic electrons. (a) Flow chart showing the three steps: (Step I) density-functional band calculation, (Step II) calculations for maximally localized Wannier function and transfer integral $t$ [Eq. (1)], and (Step III) calculations for effective on-site $U$ and nearest-neighbor $U_{nn}$ Coulomb repulsions, based on cRPA [Eq. (2)]. (b) Upper panel: A comparison between the calculated density-functional band structure (red solid curves) and the Wannier interpolated bands (blue crosses) for the anionic electrons of Na-electro-sodalite. EF represents the Fermi energy. Lower panel: Calculated real-space Wannier function (isosurface of 0.04) for the anionic electron (light blue) drawn by VESTA [38]. With the resulting values of $|t|$ = 55 meV, $U$ = 2.86 eV, and $U_{nn}$ = 0.66 eV, the correlation strength of the sodalite is calculated as 39.8.

## III. RESULTS

Before showing the calculation results for the electronic correlation strength of electrides, we describe a classification of the present electrides in terms of the dimensionality of the Wannier-function network of anionic electrons. Figure 2 shows examples of the electrides classified by the dimension of the Wannier-function-network: (a) Na-electro-sodalite (0D), (b) $Sr_5P_3$ (1D), (c) $Ca_2N$ (2D), and (d) $LaH_2$ (3D). In the 0D electrides, each Wannier function is confined in a cage and has no connection to adjacent Wannier functions. In the 1D case, the Wannier functions have connections along a particular direction. Similarly, in the 2D case, the Wannier functions form layers, and each layer is separated by a nanolayer block. In the 3D case, the Wannier functions create a 3D network. For electrides, hybridization between the Wannier orbital of the anionic electrons and the atomic orbitals of the cages plays an important role in network formation. In the case of $LaH_2$ [Fig. 2(d)], the anionic electron orbital is hybridized with the La-*d* orbital, resulting in a 3D network of Wannier functions. The dimensionality of the Wannier-function network is quite useful for understanding the low-energy properties of the electride, as explained later.

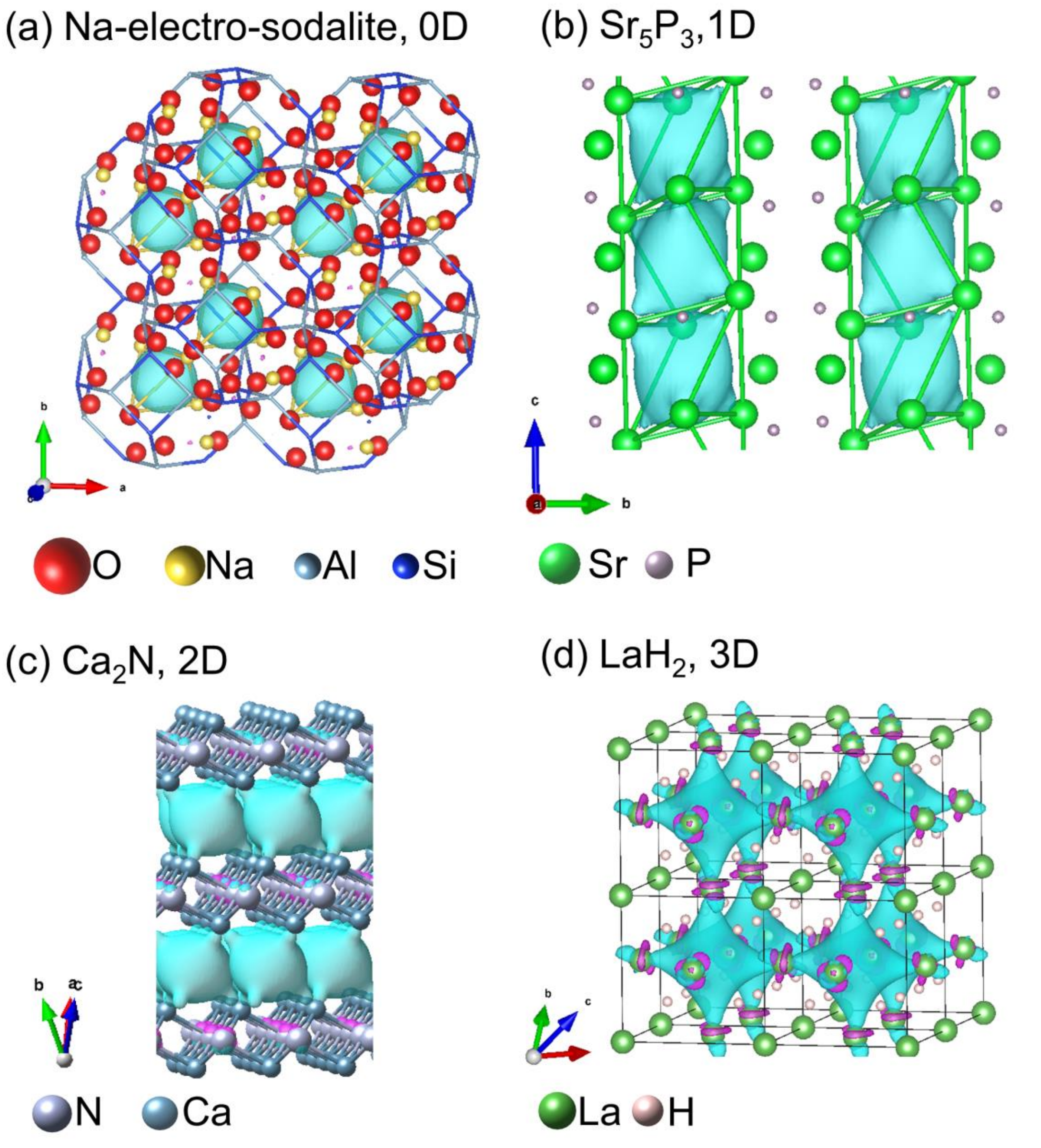

FIG. 2 Classification of Wannier-function network from dimensionality: (a) Na-electro-sodalite (0D), (b) $Sr_5P_3$ (1D), (c) $Ca_2N$ (2D), and (d) $LaH_2$ (3D). In these panels, we display the Wannier functions for multiple unit cells to show the connectivity of the Wannier functions. Note that the Wannier function itself has a localized nature in the unit cell. The isosurfaces are 0.04.

Figure 3 shows the calculated correlation strength $(U - U_{nn})/|t|$ of 19 electrides, together with the $U$, $U_{nn}$, and $|t|$ parameters. The calculated values are given in Table A2 in supplemental material A. We confirm an appreciable material dependence for the correlation strength; the differences in the Wannier-function networks bring about big differences in the correlation strength. From the figure, we find that the correlation strength varies in the order 0D >> 1D> 2D ~ 3D; the strengths of the 0D systems are larger than 10, while those of the 2D and 3D systems are smaller than 3. Thus, it is expected that the 0D electrides would exhibit strong correlation effects such as the realization of Mott insulators and magnetic states, while the 2D and 3D electrides would be metals or band insulators because the anionic electrons behave like free electrons. The observed trend of the correlation strength on the dimensionality is consistent with experiments [6,7,10,11,39].

There are also some remarkable trends. In the 1D electrides, the correlation strength of the $Ca_5Pn_3$ ($Pn$=P, As, Sb) group is smaller than that of the $Sr_5Pn_3$ group, and that of the latter group is smaller than that of the $Ba_5Pn_3$ group. For example, the values of 1D $Ca_5P_3$, $Sr_5P_3$, and $Ba_5P_3$ are 4.0, 5.7, and 9.7, respectively, and thus 1D $Ba_5Pn_3$ ($Pn$=P, As, Sb) will be of interest for future experimental studies since 1D $Sr_5P_3$ has already been shown to be a Mott insulator. The increasing trend in correlation strength, $Ca_5Pn_3$ < $Sr_5Pn_3$ < $Ba_5Pn_3$, originates from the decreasing trend in the transfer integrals $|t|$ according to $Ca_5Pn_3$ > $Sr_5Pn_3$ > $Ba_5Pn_3$ (see Fig. 3). The 1D $Ae_5Pn_3$ ($Ae$=Ca, Sr, Ba) electrides have a hexagonal structure, and the Wannier functions form a 1D network along the $c$ axis. Therefore, the transfer integrals $|t|$ depend sensitively on the $c$ parameter of the system. The $c$ parameters of $Ca_5P_3$, $Sr_5P_3$, and $Ba_5P_3$ are 6.89 Å, 7.46 Å, and 8.07 Å, respectively, and the transfer integrals are 164, 129, and 58 meV, respectively. Thus, the cation species in 1D $Ae_5Pn_3$ has a considerable effect on the strength of the electronic correlation via the transfer $|t|$.

Besides, in the 0D electrides, we identified a group having considerably small Coulomb interactions and transfer integrals: $Ae_5Sb_3$ with the $Yb_5Sb_3$ structure ($U$ = 0.76 eV, $U_{nn}$ = 0.14 eV, and $|t|$ = 21 meV on average). To understand the trend in 0D $Ae_5Sb_3$, Fig. 4 compares the band structure for 0D $Sr_5Sb_3$ [panel (a)] with that for 0D C12A7:$e^-$ [panel (b)] having considerably large values of $U$ = 2.63 eV, $U_{nn}$ = 0.89 eV, and $|t|$ = 113 meV. It can be seen that $Sr_5Sb_3$ has a fairly dense band structure in contrast to the sparse band structure of C12A7:$e^-$. Because of the dense band structure, screening due to low-energy excitation becomes effective in $Sr_5Sb_3$, so that the resulting effective Coulomb interactions are significantly reduced. In addition, the Wannier bandwidths calculated for $Sr_5Sb_3$ and C12A7:$e^-$ are 0.6 eV and 1.7 eV, respectively, leading to a small $|t|$ for $Sr_5Sb_3$. We emphasize that the 0D $Ae_5Sb_3$ group is still a strongly correlated electron system having large correlation strengths of more than 10 despite the small values of $U$, $U_{nn}$, and $|t|$.

Finally, we remark on the weak correlation trend in the 2D and 3D electrides. As can be seen from Figs. 2(c) and 2(d), the electrides have large spaces where electrons can move freely, and thus the electrides can have a large kinetic energy. Thus, the electronic correlation would not be strong in the 2D and 3D electrides. For 2D electrides, good 2D electron gas confinement can occur between the two-dimensional layers. Recently, a material with a large thermoelectric coefficient has been found in a 2D electron gas system [40], and research on such a system would be interesting for the study of electrides (the thermoelectric properties of 0D C12A7:$e^-$ have been already reported [41]). 3D electrides are actually quite similar to 0D electrides; the difference is that the 0D anionic electrons are well isolated [Fig.2(a)], whereas, in the 3D system, the anionic electrons are spatially spread through hybridization with neighboring orbitals of anionic electrons. [Fig. 2(d)]. Therefore, by changing the cation walls, it might be possible to control the electronic correlation in 0D and 3D electrides in the weak-correlation to strong-correlation regimes, which is important for verifying strong correlation phenomena such as metal-insulator transitions [42].

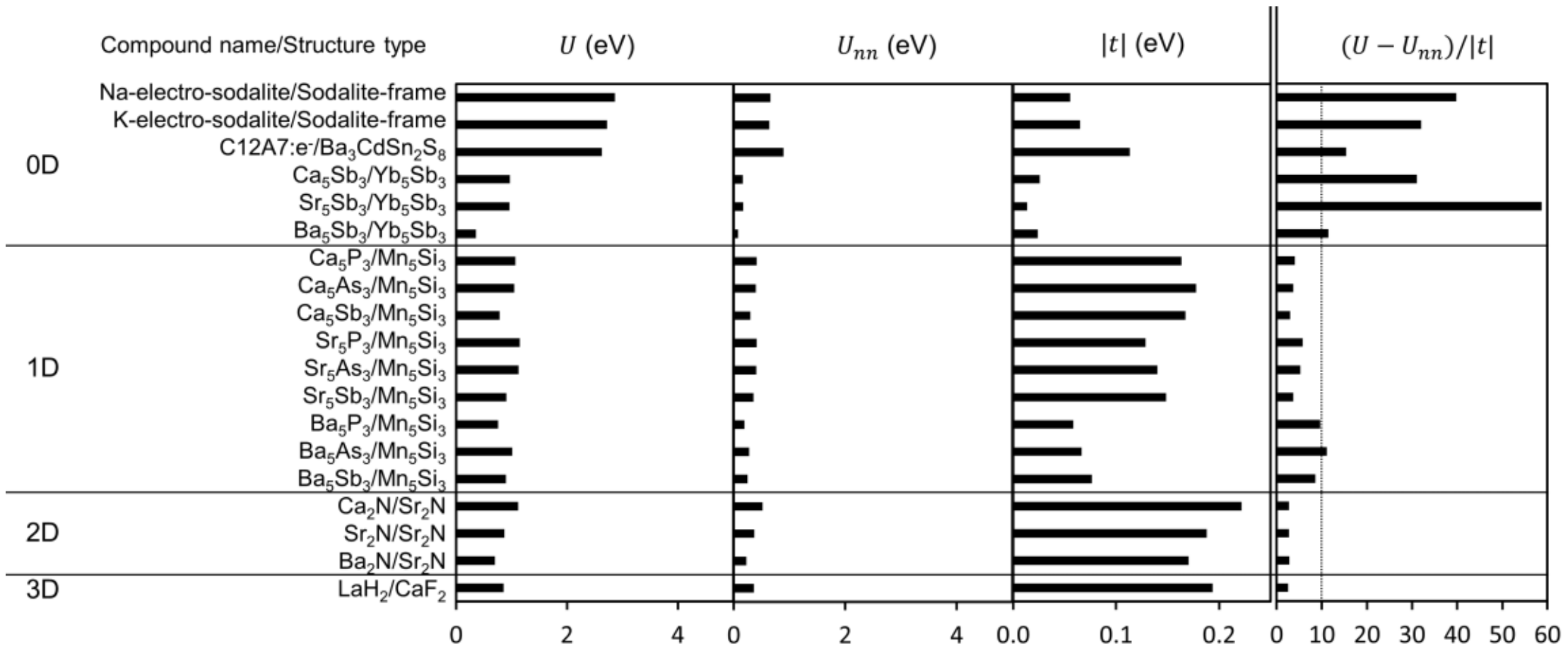

FIG. 3. Comparative chart of calculated cRPA onsite $U$ and nearest-neighbor $U_{nn}$ Coulomb interactions [Eq. (2)], nearest-neighbor transfer integral $|t|$ [Eq. (1)], and electronic correlation strength $(U - U_{nn})/|t|$ of 19 inorganic electrides including 0D, 1D, 2D, and 3D systems. The dotted line for $(U - U_{nn})/|t|$ = 10 is given as a guide, which is a rough threshold for the appearance of strong correlation phenomena. The crystal structures of all the structure types used in this study are shown in Fig. A1 and Table A3 of supplemental material A.

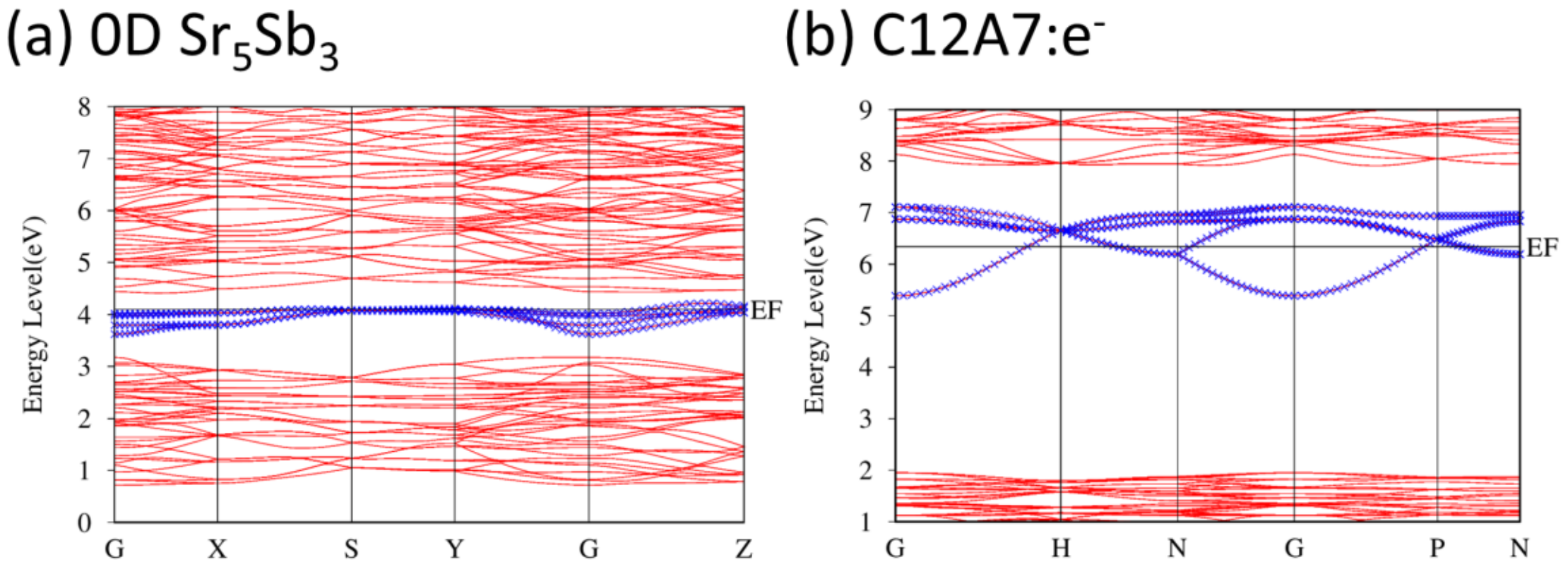

FIG. 4. Calculated density-functional band structures (red solid curves) and Wannier interpolated bands (blue crosses) for anionic electrons in (a) 0D $Sr_5Sb_3$ and (b) 0D C12A7:$e^-$. EF is the Fermi level. Band dispersions are plotted along the high symmetry points, where G = (0, 0, 0), X = ($a$*/2, 0, 0), S = ($a$*/2, $b$*/2, 0), Y = (0, $b$*/2, 0), Z = (0, 0, $c$*/2), H = (-$a$*/2, $b$*/2, $c$*/2), N = (0, 0, $c$*/2), and P = ($a$*/4, $b$*/4, $c$*/4), with $a$*, $b$*, and $c$* being the basis vectors of the reciprocal lattice.

# IV. CONCLUSION

In summary, we present a first-principles analysis of electronic correlation of inorganic electrides. In the present study, we classified the electrides from the viewpoint of the electronic structure, focusing on the Wannier-function network of anionic electrons. The network analysis is effective for obtaining a systematic understanding of the electronic correlation of electrides. We derived the transfer integral $|t|$ and effective interaction parameters $U$ and $U_{nn}$ using the maximally localized Wannier function and cRPA for 19 electrides, including 0D, 1D, 2D, and 3D systems. Through comparison of the electronic correlation strength $(U - U_{nn})/|t|$ of the electrides, we found that the 0D electrides and some of the 1D electrides exhibit strong electronic correlation, while the higher dimensional electrides basically exhibit weaker electronic correlation dominated by the gain of kinetic energy in free space. The trend of the electronic correlation on the dimensionality is consistent with experimental observations.

The present analysis proposes practical ideas for controlling the electronic properties of electrides. By controlling and designing the nanospaces in electrides, bandwidth control is possible. This is particularly important for 1D electrides, where the bandwidth depends sensitively on the type of cations. Thus, they would be expected to be prototype electrides for investigating correlation phenomena such as metal-insulator and magnetic transitions. Also, accurate filling control for the electrides is important for further study; 0D C12A7:$e^-$ has 0.7 electrons per crystalline void and 2D $Y_2C$ [9,43] has 1.8 electrons per void. They are not half-filling, and the observation of strong correlation phenomena by doping control would be expected.

# ACKNOWLEDGMENTS

This work was supported by funds from Kakenhi Grant-in-Aid (No. 17H06153) from the Japan Society for the Promotion of Science (JSPS) and the Ministry of Education, Culture, Sports, Science and Technology (MEXT) Element Strategy Initiative to form a research core (No. JPMXP0112101001). K.N. acknowledges the financial support of JSPS Kakenhi Grant No. 16H06345, No. 16K05452, No. 17H03393, No. 17H03379, and No. 19K03673.

**Supplementary Material**

# Electronic correlation strength of inorganic electrides from first principles

Shu Kanno[1], Tomofumi Tada[1,*], Takeru Utsumi[1], Kazuma Nakamura[2], Hideo Hosono[1,†]

[1]*Materials Research Center for Element Strategy, Tokyo Institute of Technology, 4259 Nagatsuta, Midori-ku, Yokohama 226-8503, Japan*

[2]*Quantum Physics Section, Department of Basic Sciences, Kyushu Institute of Technology, 1-1 Sensui-cho, Tobata, Kitakyushu, Fukuoka, 804-8550, Japan*

*** tada.t.ae@m.titech.ac.jp**

**† hosono@mces.titech.ac.jp**

## A. Computational conditions and calculated results

Table A1 lists the detailed computational conditions, and Table A2 the calculated results for $U$, $U_{nn}$, $|t|$, and $(U - U_{nn})/|t|$, where $U$ is the effective on-site Coulomb repulsion, $U_{nn}$ is the effective nearest-neighbor Coulomb repulsion, and $|t|$ is the absolute value of the nearest-neighbor transfer integral. We obtained the crystal structural data by using the Materials Project [1] database and Inorganic Crystal Structure Database (ICSD) [2], as listed in the column "Reference" in Tables A1 and A2. When the column is blank, the structures of the systems were prepared based on an experimentally confirmed compound in the same structure type. Figure A1 shows the crystal structures for all the structure types used in this study, and Table A3 shows the lattice and structure parameters of the crystal structures.

TABLE A1

The computational conditions for all the compounds calculated in this study

| Dimension | Compound | Structure type | Wave function cut-off (eV) | *k*-point grid | Number of bands | Dielectric function cutoff (eV) | Reference |
|---|---|---|---|---|---|---|---|
| 0D | Na-electro-sodalite | Sodalite-frame | 81 | 7x7x7 | 300 | 3 | [1] |
| 0D | K-electro-sodalite | Sodalite-frame | 49 | 5x5x5 | 200 | 5 | [2] |
| 0D | C12A7:$e^-$ | $Ba_3CdSn_2S_8$ | 81 | 6x6x6 | 350 | 3 | [1] |
| 0D | $Ca_5Sb_3$ | $Yb_5Sb_3$ | 81 | 5x5x5 | 240 | 3 | [2] |
| 0D | $Sr_5Sb_3$ | $Yb_5Sb_3$ | 81 | 5x5x5 | 240 | 3 | [2] |
| 0D | $Ba_5Sb_3$ | $Yb_5Sb_3$ | 81 | 5x5x5 | 240 | 3 | |
| 1D | $Ca_5P_3$ | $Mn_5Si_3$ | 81 | 5x5x5 | 120 | 3 | |
| 1D | $Ca_5As_3$ | $Mn_5Si_3$ | 81 | 5x5x5 | 120 | 3 | [2] |
| 1D | $Ca_5Sb_3$ | $Mn_5Si_3$ | 81 | 5x5x5 | 160 | 3 | [2] |
| 1D | $Sr_5P_3$ | $Mn_5Si_3$ | 50 | 7x7x9 | 120 | 5 | [3] |
| 1D | $Sr_5As_3$ | $Mn_5Si_3$ | 81 | 5x5x5 | 160 | 3 | [2] |
| 1D | $Sr_5Sb_3$ | $Yb_5Sb_3$ | 81 | 5x5x5 | 240 | 3 | |
| 1D | $Ba_5P_3$ | $Mn_5Si_3$ | 81 | 5x5x5 | 160 | 3 | |
| 1D | $Ba_5As_3$ | $Mn_5Si_3$ | 81 | 5x5x5 | 160 | 3 | [2] |
| 1D | $Ba_5Sb_3$ | $Yb_5Sb_3$ | 81 | 5x5x5 | 240 | 3 | [2] |
| 2D | $Ca_2N$ | $Sr_2N$ | 81 | 8x8x8 | 60 | 1.5 | [1] |
| 2D | $Sr_2N$ | $Sr_2N$ | 81 | 8x8x8 | 60 | 3 | [1] |
| 2D | $Ba_2N$ | $Sr_2N$ | 81 | 8x8x8 | 60 | 3 | [1] |
| 3D | $LaH_2$ | $CaF_2$ | 81 | 11x11x11 | 50 | 3 | [1] |

TABLE A2

Calculated $U$, $U_{nn}$, $|t|$, and $(U - U_{nn})/|t|$ of all the compounds calculated in this study

| Dimension | Compound | Structure type | $U$(eV) | $U_{nn}$(eV) | $\lvert t\rvert$(eV) | $\frac{U-U_{nn}}{\lvert t\rvert}$ | Reference |
|---|---|---|---|---|---|---|---|
| 0D | Na-electro-sodalite | Sodalite-frame | 2.86 | 0.66 | 0.055 | 39.8 | [1] |
| 0D | K-electro-sodalite | Sodalite-frame | 2.72 | 0.64 | 0.065 | 32.0 | [2] |
| 0D | C12A7:e$^-$ | $Ba_3CdSn_2S_8$ | 2.63 | 0.89 | 0.113 | 15.4 | [1] |
| 0D | $Ca_5Sb_3$ | $Yb_5Sb_3$ | 0.97 | 0.17 | 0.026 | 31.1 | [2] |
| 0D | $Sr_5Sb_3$ | $Yb_5Sb_3$ | 0.96 | 0.17 | 0.014 | 58.7 | [2] |
| 0D | $Ba_5Sb_3$ | $Yb_5Sb_3$ | 0.35 | 0.08 | 0.024 | 11.5 | |
| 1D | $Ca_5P_3$ | $Mn_5Si_3$ | 1.07 | 0.41 | 0.164 | 4.0 | |
| 1D | $Ca_5As_3$ | $Mn_5Si_3$ | 1.05 | 0.40 | 0.178 | 3.7 | [2] |
| 1D | $Ca_5Sb_3$ | $Mn_5Si_3$ | 0.79 | 0.30 | 0.167 | 2.9 | [2] |
| 1D | $Sr_5P_3$ | $Mn_5Si_3$ | 1.15 | 0.41 | 0.129 | 5.7 | [3] |
| 1D | $Sr_5As_3$ | $Mn_5Si_3$ | 1.13 | 0.40 | 0.140 | 5.2 | [2] |
| 1D | $Sr_5Sb_3$ | $Mn_5Si_3$ | 0.90 | 0.35 | 0.149 | 3.7 | |
| 1D | $Ba_5P_3$ | $Mn_5Si_3$ | 0.76 | 0.19 | 0.058 | 9.7 | |
| 1D | $Ba_5As_3$ | $Mn_5Si_3$ | 1.01 | 0.27 | 0.067 | 11.1 | [2] |
| 1D | $Ba_5Sb_3$ | $Mn_5Si_3$ | 0.90 | 0.25 | 0.076 | 8.5 | [2] |
| 2D | $Ca_2N$ | $Sr_2N$ | 1.12 | 0.51 | 0.222 | 2.7 | [1] |
| 2D | $Sr_2N$ | $Sr_2N$ | 0.87 | 0.37 | 0.188 | 2.7 | [1] |
| 2D | $Ba_2N$ | $Sr_2N$ | 0.70 | 0.22 | 0.170 | 2.8 | [1] |
| 3D | $LaH_2$ | $CaF_2$ | 0.85 | 0.36 | 0.194 | 2.5 | [1] |

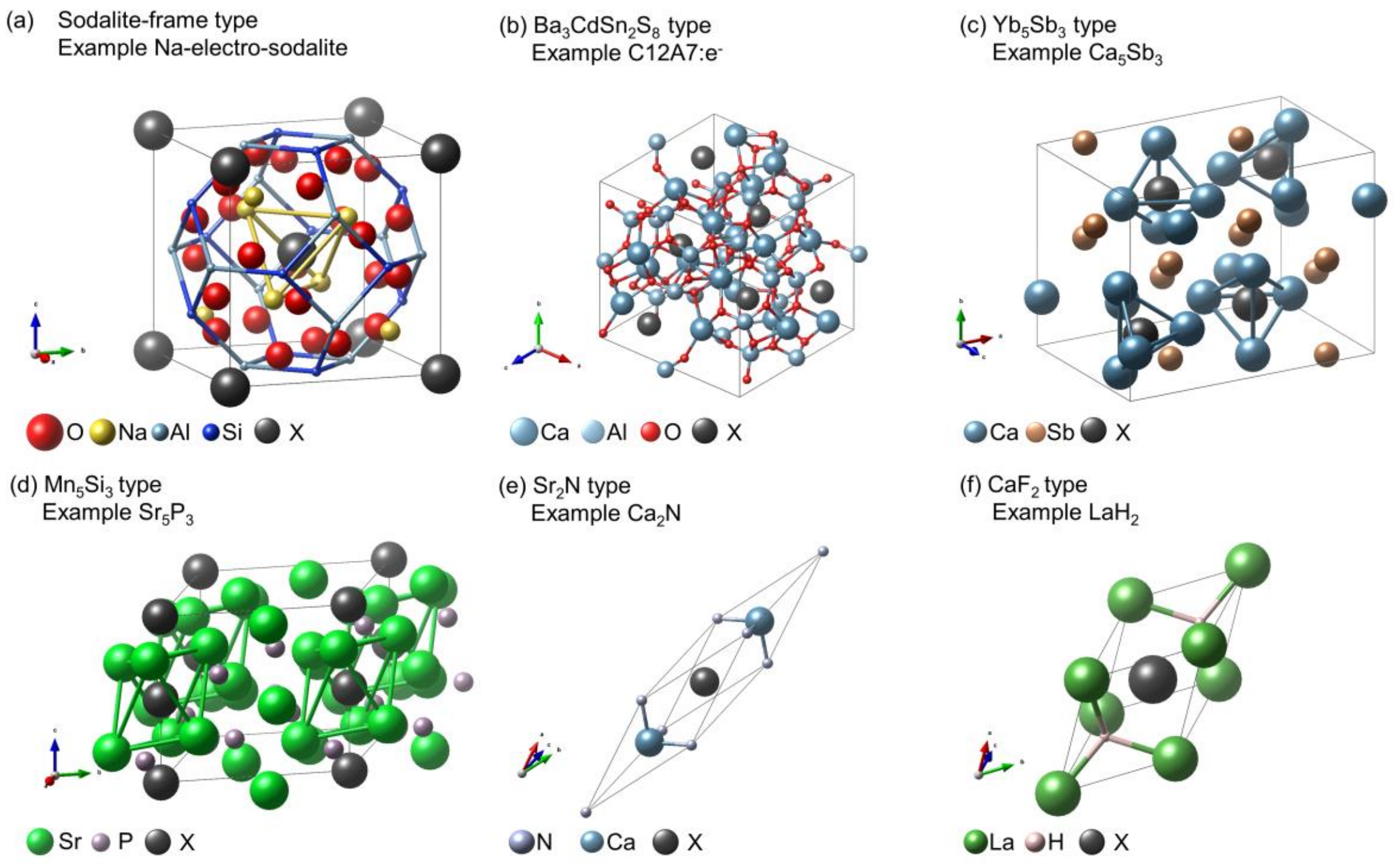

FIG. A1. Crystal structures of all the structure types used in this study. “X” represents the central position of the Wannier function; the positions can be changed slightly during the Wannier function optimization.

TABLE A3. Lattice and structure parameters of the crystal structures of all the structure types used in this study. We show the parameters for the examples of the compounds corresponding to each structure type. The fractional coordinate of “X” represents the central position of the Wannier function for the initial guess in the optimization.

| Structure type | Dimension | Example | Lattice parameter | Structure parameter | | | |
|---|---|---|---|---|---|---|---|
| | | | | Atom | Fractional coordinate | | |
| | | | | | x | y | z |
| Sodalite-frame | 0D | Na-electro-sodalite | a=8.943Å | Al | 0.250 | 0.000 | 0.500 |
| | | | b=8.943Å | Al | 0.750 | 0.000 | 0.500 |
| | | | c=8.943Å | Al | 0.500 | 0.250 | 0.000 |
| | | | α=90° | Al | 0.500 | 0.750 | 0.000 |
| | | | β=90° | Al | 0.000 | 0.500 | 0.250 |
| | | | γ=90° | Al | 0.000 | 0.500 | 0.750 |
| | | | | Si | 0.250 | 0.500 | 0.000 |
| | | | | Si | 0.750 | 0.500 | 0.000 |
| | | | | Si | 0.000 | 0.250 | 0.500 |
| | | | | Si | 0.000 | 0.750 | 0.500 |
| | | | | Si | 0.500 | 0.000 | 0.250 |
| | | | | Si | 0.500 | 0.000 | 0.750 |
| | | | | Na | 0.176 | 0.176 | 0.176 |
| | | | | Na | 0.824 | 0.824 | 0.176 |
| | | | | Na | 0.824 | 0.176 | 0.824 |
| | | | | Na | 0.176 | 0.824 | 0.824 |
| | | | | Na | 0.676 | 0.676 | 0.676 |
| | | | | Na | 0.324 | 0.324 | 0.676 |
| | | | | Na | 0.676 | 0.324 | 0.324 |
| | | | | Na | 0.324 | 0.676 | 0.324 |
| | | | | O | 0.152 | 0.435 | 0.140 |
| | | | | O | 0.849 | 0.565 | 0.140 |
| | | | | O | 0.849 | 0.435 | 0.860 |
| | | | | O | 0.152 | 0.565 | 0.860 |
| | | | | O | 0.140 | 0.152 | 0.435 |
| | | | | O | 0.140 | 0.849 | 0.565 |
| | | | | O | 0.860 | 0.849 | 0.435 |
| | | | | O | 0.860 | 0.152 | 0.565 |
| | | | | O | 0.435 | 0.140 | 0.152 |
| | | | | O | 0.565 | 0.140 | 0.849 |
| | | | | O | 0.435 | 0.860 | 0.849 |
| | | | | O | 0.565 | 0.860 | 0.152 |
| | | | | O | 0.935 | 0.652 | 0.640 |
| | | | | O | 0.065 | 0.349 | 0.640 |
| | | | | O | 0.935 | 0.349 | 0.360 |
| | | | | O | 0.065 | 0.652 | 0.360 |
| | | | | O | 0.652 | 0.640 | 0.935 |
| | | | | O | 0.349 | 0.640 | 0.065 |
| | | | | O | 0.349 | 0.360 | 0.935 |
| | | | | O | 0.652 | 0.360 | 0.065 |
| | | | | O | 0.640 | 0.935 | 0.652 |
| | | | | O | 0.640 | 0.065 | 0.349 |
| | | | | O | 0.360 | 0.935 | 0.349 |
| | | | | O | 0.360 | 0.065 | 0.652 |
| | | | | X | 0.000 | 0.000 | 0.000 |
| | | | | X | 0.500 | 0.500 | 0.500 |
| $Ba_3CdSn_2S_8$ | 0D | C12A7:$e^-$ | a=10.479Å | Ca | 0.640 | 0.250 | 0.890 |
| | | | b=10.479Å | Ca | 0.860 | 0.750 | 0.610 |
| | | | c=10.479Å | Ca | 0.390 | 0.750 | 0.140 |
| | | | α=109.471° | Ca | 0.110 | 0.250 | 0.360 |
| | | | β=109.471° | Ca | 0.890 | 0.640 | 0.250 |
| | | | γ=109.471° | Ca | 0.610 | 0.860 | 0.750 |
| | | | | Ca | 0.140 | 0.390 | 0.750 |
| | | | | Ca | 0.360 | 0.110 | 0.250 |
| | | | | Ca | 0.250 | 0.890 | 0.640 |
| | | | | Ca | 0.750 | 0.610 | 0.860 |
| | | | | Ca | 0.750 | 0.140 | 0.390 |
| | | | | Ca | 0.250 | 0.360 | 0.110 |
| | | | | Al | 0.375 | 0.250 | 0.625 |
| | | | | Al | 0.125 | 0.750 | 0.875 |
| | | | | Al | 0.625 | 0.375 | 0.250 |
| | | | | Al | 0.875 | 0.125 | 0.750 |
| | | | | Al | 0.250 | 0.625 | 0.375 |
| | | | | Al | 0.750 | 0.875 | 0.125 |
| | | | | Al | 0.500 | 0.000 | 0.037 |
| | | | | Al | 0.037 | 0.500 | 0.000 |
| | | | | Al | 0.000 | 0.037 | 0.500 |
| | | | | Al | 0.464 | 0.464 | 0.464 |
| | | | | Al | 0.537 | 0.500 | 0.000 |
| | | | | Al | 0.500 | 0.000 | 0.537 |
| | | | | Al | 0.000 | 0.537 | 0.500 |
| | | | | Al | 0.964 | 0.964 | 0.964 |
| | | | | O | 0.500 | 0.000 | 0.871 |
| | | | | O | 0.871 | 0.500 | 0.000 |
| | | | | O | 0.000 | 0.871 | 0.500 |
| | | | | O | 0.629 | 0.629 | 0.629 |
| | | | | O | 0.371 | 0.500 | 0.000 |
| | | | | O | 0.500 | 0.000 | 0.371 |
| | | | | O | 0.000 | 0.371 | 0.500 |
| | | | | O | 0.129 | 0.129 | 0.129 |
| | | | | O | 0.408 | 0.094 | 0.617 |
| | | | | O | 0.523 | 0.406 | 0.814 |
| | | | | O | 0.092 | 0.709 | 0.686 |
| | | | | O | 0.977 | 0.791 | 0.883 |
| | | | | O | 0.117 | 0.594 | 0.908 |
| | | | | O | 0.314 | 0.906 | 0.023 |
| | | | | O | 0.186 | 0.209 | 0.592 |
| | | | | O | 0.383 | 0.291 | 0.477 |
| | | | | O | 0.617 | 0.408 | 0.094 |
| | | | | O | 0.814 | 0.523 | 0.406 |
| | | | | O | 0.686 | 0.092 | 0.709 |
| | | | | O | 0.883 | 0.977 | 0.791 |
| | | | | O | 0.908 | 0.117 | 0.594 |
| | | | | O | 0.023 | 0.314 | 0.906 |
| | | | | O | 0.592 | 0.186 | 0.209 |
| | | | | O | 0.477 | 0.383 | 0.291 |
| | | | | O | 0.094 | 0.617 | 0.408 |
| | | | | O | 0.406 | 0.814 | 0.523 |
| | | | | O | 0.709 | 0.686 | 0.092 |
| | | | | O | 0.791 | 0.883 | 0.977 |
| | | | | O | 0.594 | 0.908 | 0.117 |
| | | | | O | 0.906 | 0.023 | 0.314 |
| | | | | O | 0.209 | 0.592 | 0.186 |
| | | | | O | 0.291 | 0.477 | 0.383 |
| | | | | X | 0.625 | 0.750 | 0.375 |
| | | | | X | 0.750 | 0.375 | 0.625 |
| | | | | X | 0.375 | 0.625 | 0.750 |
| | | | | X | 0.875 | 0.250 | 0.125 |
| | | | | X | 0.125 | 0.875 | 0.250 |
| | | | | X | 0.250 | 0.125 | 0.875 |
| $Yb_5Sb_3$ | 0D | $Ca_5Sb_3$ | a=12.502Å | Ca | 0.075 | 0.043 | 0.693 |
| | | | b=9.512Å | Ca | 0.925 | 0.958 | 0.307 |
| | | | c=8.287Å | Ca | 0.425 | 0.958 | 0.193 |
| | | | α=90° | Ca | 0.575 | 0.043 | 0.807 |
| | | | β=90° | Ca | 0.925 | 0.543 | 0.307 |
| | | | γ=90° | Ca | 0.075 | 0.458 | 0.693 |
| | | | | Ca | 0.575 | 0.458 | 0.807 |
| | | | | Ca | 0.425 | 0.543 | 0.193 |
| | | | | Ca | 0.228 | 0.250 | 0.321 |
| | | | | Ca | 0.773 | 0.750 | 0.679 |
| | | | | Ca | 0.273 | 0.750 | 0.821 |
| | | | | Ca | 0.728 | 0.250 | 0.179 |
| | | | | Ca | 0.289 | 0.250 | 0.852 |
| | | | | Ca | 0.711 | 0.750 | 0.149 |
| | | | | Ca | 0.211 | 0.750 | 0.352 |
| | | | | Ca | 0.789 | 0.250 | 0.649 |
| | | | | Ca | 0.506 | 0.250 | 0.465 |
| | | | | Ca | 0.494 | 0.750 | 0.536 |
| | | | | Ca | 0.994 | 0.750 | 0.965 |
| | | | | Ca | 0.006 | 0.250 | 0.036 |
| | | | | Sb | 0.170 | 0.985 | 0.067 |
| | | | | Sb | 0.830 | 0.015 | 0.933 |
| | | | | Sb | 0.330 | 0.015 | 0.567 |
| | | | | Sb | 0.670 | 0.985 | 0.433 |
| | | | | Sb | 0.830 | 0.485 | 0.933 |
| | | | | Sb | 0.170 | 0.515 | 0.067 |
| | | | | Sb | 0.670 | 0.515 | 0.433 |
| | | | | Sb | 0.330 | 0.485 | 0.567 |
| | | | | Sb | 0.983 | 0.250 | 0.422 |
| | | | | Sb | 0.017 | 0.750 | 0.578 |
| | | | | Sb | 0.517 | 0.750 | 0.922 |
| | | | | Sb | 0.483 | 0.250 | 0.078 |
| | | | | X | 0.112 | 0.250 | 0.818 |
| | | | | X | 0.612 | 0.250 | 0.682 |
| | | | | X | 0.388 | 0.750 | 0.318 |
| | | | | X | 0.888 | 0.750 | 0.182 |
| $Mn_5Si_3$ | 1D | $Sr_5P_3$ | a=8.696Å | Sr | 0.741 | 0.998 | 0.199 |
| | | | b=8.696Å | Sr | 0.002 | 0.259 | 0.801 |
| | | | c=7.464Å | Sr | 0.259 | 0.002 | 0.801 |
| | | | α=90.679° | Sr | 0.998 | 0.741 | 0.199 |
| | | | β=90.679° | Sr | 0.351 | 0.649 | 0.000 |
| | | | γ=119.480° | Sr | 0.649 | 0.351 | 0.000 |
| | | | | Sr | 0.685 | 0.315 | 0.500 |
| | | | | Sr | 0.315 | 0.685 | 0.500 |
| | | | | Sr | 0.255 | 0.255 | 0.331 |
| | | | | Sr | 0.745 | 0.745 | 0.669 |
| | | | | P | 0.609 | 0.999 | 0.767 |
| | | | | P | 0.001 | 0.391 | 0.233 |
| | | | | P | 0.391 | 0.001 | 0.233 |
| | | | | P | 0.999 | 0.609 | 0.767 |
| | | | | P | 0.393 | 0.393 | 0.728 |
| | | | | P | 0.607 | 0.607 | 0.272 |
| | | | | X | 0.000 | 0.000 | 0.000 |
| | | | | X | 0.000 | 0.000 | 0.500 |
| $Sr_2N$ | 2D | $Ca_2N$ | a=6.747Å | Ca | 0.268 | 0.268 | 0.268 |
| | | | b=6.747Å | Ca | 0.732 | 0.732 | 0.732 |
| | | | c=6.747Å | N | 0.000 | 0.000 | 0.000 |
| | | | α=31.029° | X | 0.500 | 0.500 | 0.500 |
| | | | β=31.029° | | | | |
| | | | γ=31.029° | | | | |
| $CaF_2$ | 3D | $LaH_2$ | a=3.999Å | La | 0.000 | 0.000 | 0.000 |
| | | | b=3.999Å | H | 0.750 | 0.750 | 0.750 |
| | | | c=3.999Å | H | 0.250 | 0.250 | 0.250 |
| | | | α=60° | X | 0.500 | 0.500 | 0.500 |
| | | | β=60° | | | | |
| | | | γ=60° | | | | |